\documentclass[aps,prc,onecolumn,superscriptaddress,preprint]{revtex4}
\usepackage{graphicx}
\boldmath
\usepackage{epsfig,graphics}
\usepackage{graphicx}
\usepackage{dcolumn}
\usepackage{amsmath}
\usepackage[usenames]{color}
\usepackage{ulem} 
\usepackage{SIunits}
\usepackage{multirow}

\begin{document}

\title{Symmetry energy and nucleon-nucleon cross sections}

\author{M.~Veselsk\'{y}}
\affiliation{Institute of Physics, Slovak Academy of Sciences, D\'{u}bravsk\'{a} cesta 9, 84511 Bratislava}
\email[]{e-mail: martin.veselsky@savba.sk}

\author{Y. G. Ma}
\affiliation{Shanghai Institute of Applied Physics, Chinese Academy of Sciences,
2019 Jia-Luo Road,  P.O. Box 800-204, 
Shanghai 201800, China}
\email[]{e-mail: ygma@sinap.ac.cn}

\date{\today}



\begin{abstract}
The extension of the Boltzmann-Uehling-Uhlenbeck model of nucleus-nucleus collision is presented. 
The isospin-dependent 
nucleon-nucleon cross sections are estimated using the proper volume extracted from the equation 
of state of the nuclear matter transformed into the form of the Van der Waals equation of state.  
The results of such simulations demonstrate  
the dependence on symmetry energy which typically varies  
strongly from the results obtained using only the isospin-dependent mean-field. 
The evolution of the n/p multiplicity ratio with angle and kinetic energy, 
in combination with the elliptic flow of neutrons and protons, provides 
a suitable set of observables for determination of the density dependence 
of the symmetry energy. 
The model thus provides an environment for testing of equations of state, 
used for various applications in nuclear physics and astrophysics. 
\end{abstract}

\pacs{
pacs
}

\maketitle

\section*{Introduction}
One of the main goals of intermediate-energy heavy-ion
collisions (HIC) is to study properties of nuclear matter,
especially to determine the nuclear equation of state (EoS).
HIC provide a unique possibility to compress nuclear matter to
a hot and dense phase within a laboratory environment. The
pressures that result from the high densities achieved during
such collisions strongly influence the motion of ejected matter
and are sensitive to the EoS.  Within the hard work of the researchers over the three decades, the 
EoS of symmetric nuclear matter was studied in detail by the study of giant dipole resonances, collective flow as well as 
multifragmentation \cite{Youn99,Nato02,Dani02,Reis12}. The EoS of  isospin asymmetric nuclear matter is recently underway, 
particularly, for the density dependence of symmetry energy. 
Considerable progress has been made in determining the sub- and supra-saturation density behavior of the symmetry energy 
\cite{Tsan04,Fami06,Tsan09,Kuma11a,Xiao09,Rus11,Cozm11,Gaut11}. The later part is still an 
unanswered question in spite of recent findings in term of neutron-proton 
elliptic  flow ratio and difference \cite{Rus11,Cozm11}. However, the  former one is understood to 
some extent \cite{Tsan04,Fami06,Tsan09,Kuma11a}, although, more efforts are needed for precise measurements. 

Transport model is very useful to treat heavy ion collision
dynamics and obtain important information of nuclear matter EoS as well as the symmetry energy.  In
intermediate energy heavy ion collisions, the
Boltzmann-Uehling-Uhlenbeck model is an  extensively useful tool
\cite{BUU1,BUU2}, which takes both Pauli blocking and mean field into
consideration. The BUU equation reads

\begin{align}
      & \frac{\partial f}{\partial t}+ v \cdot \nabla_r f - \nabla_r U
\cdot \nabla_p f  = \frac{4}{(2\pi)^3} \int d^3p_2 d^3p_3 d\Omega
\nonumber
\\ & \frac{d\sigma_{NN}}{d\Omega}v_{12}
 \times [f_3 f_4(1-f)(1-f_2) - f f_2(1-f_3)(1-f_4)] \nonumber
\\ & \delta^3(p+p_2-p_3-p_4),  \label{BUU}
                   \end{align}
where  $f$=$f(r,p,t)$ is the phase-space distribution function.
 It is solved with the test particle method of Wong 
\cite{Wong}, with the collision term as introduced  
by Cugnon, Mizutani and Vandermeulen \cite{Cugnon}. 
In Eq.(~\ref{BUU}), $\frac{d\sigma_{NN}}{d\Omega}$
and $v_{12}$ are in-medium nucleon-nucleon cross section and
relative velocity for the colliding nucleons, respectively, and
$U$ is the single-particle mean field potential with the addition 
of the isospin-dependent symmetry energy term: 

\begin{equation}
{U} =  a\rho + b\rho^{\kappa}
+ 2{a_s}(\frac{\rho}{\rho_0})^{\gamma} \tau_z I , 
\label{EqPot}
\end{equation}
where 
$I=(\rho_n-\rho_p)/\rho$, 
$\rho_0$ is the normal nuclear matter density; $\rho$,
$\rho_n$, and $\rho_p$ are the nucleon, neutron and proton
densities, respectively; 
$\tau_z$ assumes value 
1 for neutron and -1 for proton,  
coefficients $a$, $b$ and $\kappa$ represent properties of the 
symmetric nuclear matter while the last term, which describes the influence 
of the symmetry energy,  
can be obtained e.g. from simple Weizsacker formula, where $a_s$ represents 
the coefficient of the symmetry energy term 
and $\gamma$ is the exponent, describing the density dependence. 
Typical sets of mean
field parameters cover substantial range, between the soft EoS with the
compressibility $K$ of 200 MeV ($\kappa$ = 7/6$, a \rho_0$= -356 MeV, $b \rho_0^{\kappa}$ = 303 MeV), and the hard EoS with
$K$ of 380 MeV ($\kappa$ = 2, $a \rho_0$ = -124 MeV, $b \rho_0^{\kappa}$ = 70.5 MeV) \cite{BUU1}.


It is the aim of the present work to estimate the effect of the symmetry energy parametrization within the equation of the state on the crucial component of the transport simulations, namely the in-medium nucleon-nucleon cross section.

\section*{Isospin-dependent nucleon-nucleon cross sections}

When considering influence of the symmetry energy on emission 
rates of nucleons in nucleus-nucleus collisions, one needs to understand 
whether and how the medium represented by the equation of state can influence 
relative probabilities of emission of protons and neutrons. 
Theoretical investigations of the density-dependence of in-medium nucleon-nucleon cross section were carried out for symmetric nuclear matter \cite{Machleidt,Alm}, and significant influence of nuclear density on resulting in-medium cross sections was observed in their density, angular  and energy dependencies. Using momentum-dependent interaction, ratios of in-medium to free nucleon-nucleon cross sections were evaluated via reduced nucleonic masses \cite{LiBAnn} and used for transport simulations. Still, transport simulation are mostly performed using parametrizations of the free nucleon-nucleon cross sections, eventually scaling them down empirically or using simple prescriptions for density-dependence of the scaling factor \cite{Klak}. In the present work, a prescription for estimation of the density-dependence of the in-medium nucleon-nucleon cross sections corresponding to the specific form of phenomenological nuclear equation of state will be presented.  
Such possibility to establish a simple dependence of nucleon-nucleon cross 
sections on density, temperature and symmetry energy is potentially important 
for a wide range of problems in nuclear physics and astrophysics. 

\subsection*{Equation of state of nucleonic matter}
 
Based on the single-particle potential, shown in Eq. (\ref{EqPot}), one can construct corresponding equation of the state. 
Change of the pressure in thermodynamical equation of state, 
which is also a measure of non-ideality of a neutron 
or a proton gas, can be evaluated as

\begin{equation}
\Delta p_{non-ideal} = - \frac{d\cal{U}}{dV}|_{T=const} 
\end{equation}

where $\cal{U}$ is the thermodynamic potential, $V$ is the volume and $T$ is the temperature.  
When evaluating the thermodynamic potential $\cal{U}$ as a sum of single-particle contributions of neutrons and protons, given by equation (\ref{EqPot}), 
one arrives to expression 

\begin{eqnarray}
p = x_{n}(\frac{f_{5/2}(z_n)}{f_{3/2}(z_n)} \rho T + a\rho^2 + b \kappa \rho^{1+\kappa}
+ {2 \gamma a_s \rho_0}(\frac{\rho}{\rho_0})^{1+\gamma} {I}) \nonumber \\
+ \hbox{ } x_{p} (\frac{f_{5/2}(z_p)}{f_{3/2}(z_p)} \rho T + a\rho^2 + b \kappa \rho^{1+\kappa}
- {2 \gamma a_s \rho_0}(\frac{\rho}{\rho_0})^{1+\gamma} {I}) .
\label{NuclEoS1}
\end{eqnarray}

where $x_n = \rho_n /\rho$ and $x_p = \rho_p /\rho$ are neutron and proton concentrations and 
$\frac{f_{5/2}(z)}{f_{3/2}(z)}$ 
is the factor, a fraction of the Fermi integrals ${f_{n}(z)}$, assuring that Fermi statistics is taken into 
account. The parameters $z_n = \mu_n/T$, $z_p = \mu_p/T$ are the values of fugacity of neutrons and protons, with $\mu_n$, $\mu_p$ being the neutron and proton chemical potential, respectively. 
This expression appears to provide separate 
terms for pressure of neutrons and protons, which however can be combined to obtain the typical 
quadratic dependence on isospin asymmetry $I$. The resulting pressure is the weighted average between two terms, 
which can be, in similar manner to equation (\ref{EqPot}), summarily expressed as   

\begin{equation}
p = \langle\frac{f_{5/2}(z)}{f_{3/2}(z)}\rangle \rho T + a\rho^2 + b \kappa \rho^{1+\kappa}
+ {2 \gamma a_s \rho_0}(\frac{\rho}{\rho_0})^{1+\gamma}\tau_z {I} .
\label{NuclEoS}
\end{equation}

These terms can be interpreted as the equation of state of the system of particles with the corresponding single-particle potential, given by the equation (\ref{EqPot}).  

\subsection*{Proper volume in the Van der Waals equation of state}
 
In order to find relation between equation of state and nucleon emission rates 
one can turn attention specifically to the Van der Waals equation of state. 
It can be written, using particle density $\rho$,  

\begin{equation}
(p+a'\rho^2)(1-\rho b') = \langle\frac{f_{5/2}(z)}{f_{3/2}(z)}\rangle \rho T,
\end{equation}

where the parameter $a'$ is related to attractive interaction among 
particles and $b'$ represents the proper volume of the constituent particles. 
In geometrical picture the proper volume of the particle can be directly 
related to its cross section for interaction with other particles. 
It is possible to formally transform the above equations of state for neutrons and protons (\ref{NuclEoS}) (and practically any other equation of state) 
into the form analogous to the Van der Waals equation. Then, by comparison, one obtains values of coefficients 

\begin{equation}
a' = - a,
\end{equation}

and 

\begin{equation}
b' = \frac {b\kappa \rho^{\kappa} 
+ {2 \gamma a_s}(\frac{\rho}{\rho_0})^{\gamma}\tau_z {I}} 
{p - a\rho^2}  
= \frac {b\kappa \rho^{\kappa} 
+ {2 \gamma a_s}(\frac{\rho}{\rho_0})^{\gamma}\tau_z {I}} 
{ 
\langle\frac{f_{5/2}(z)}{f_{3/2}(z)}\rangle \rho T + b\kappa \rho^{1+\kappa} + 
{2 \gamma \rho_0 a_s}(\frac{\rho}{\rho_0})^{1+\gamma}\tau_z {I}} ,
\label{bfermi}
\end{equation}

where the latter provides a measure of the proper volume of the constituents, nucleons in this case, 
as a measure 
of deviation from the behavior of the ideal gas. The proper volume 
of nucleon can be used to estimate its cross section  within the 
nucleonic medium

\begin{equation}
\sigma = (\frac{9 \pi}{16})^{1/3}~b'^{2/3}, 
\label{csvdw}
\end{equation}
which can be implemented into the collision term of the Boltzmann equation. 

Concerning the physical meaning of this procedure, for each point in the 
$\rho$-T plane the Van der Waals equation of state is found which behaves identically to the nuclear equation of state (\ref{NuclEoS}) in the vicinity of that point. Thus the dynamics of the system can be described using the two parameters of the Van der Waals EoS, of which one provides a measure of the effective volume of the constituent at a given density and temperature. The variation of the constituent volume reflects the interplay of the long-range attractive interaction, leading to its apparent increase, with the short-range repulsive interaction, leading to its apparent decrease.   

\begin{figure}[h]
\centering
\vspace{5mm}
\includegraphics[width=11.5cm,height=10.75cm]{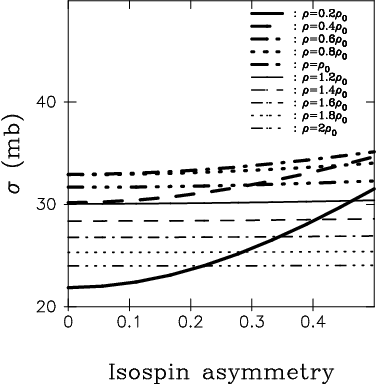}
\caption{
Extracted isospin dependencies of nucleon-nucleon cross sections for  temperature 20 MeV and various densities.
}
\label{fgcsidep}
\end{figure}

The trend of the estimated values of nucleus-nucleus cross sections,  
obtained using the soft parametrization of symmetry energy with 
$\rho^{2/3}$-dependence is shown in the Figure \ref{fgcsidep} for the temperature 20 MeV.  
The extracted values of the cross sections 
are surprisingly close to the expected value of geometric cross sections. 
With increasing isoscalar density the values of cross sections 
initially grow until they reach maximum values in the region around 
half of saturation density and then monotonously decrease. 
The increase at low densities possibly describes gradual deviation 
from the equation of state of ideal gas due to increasing attractive potential 
while at higher densities the decreasing trend may indeed represent the properties of short-range 
repulsive interaction. 
At high densities, with increasing isoscalar density 
the sensitivity to symmetry energy tends to decrease and around 
the density 2$\rho_0$ it is practically lost, which however can be preserved 
using harder parametrizations of the symmetry energy. 

It is worth mentioning that a similar rise and fall of nucleon-nucleon cross sections 
was observed by the Alm, Ropke and Schmidt \cite{Alm} and it was explained as a precursor effect of super-fluid phase-transition. Also in the present case this effect can be related to the 
phase transition in the nuclear matter, since it is caused by the same interplay of attractive and repulsive interaction which influences also the proper volume and thus extracted cross sections.  

\begin{figure}[h]
\centering
\vspace{5mm}
\includegraphics[width=8.5cm,height=8.25cm]{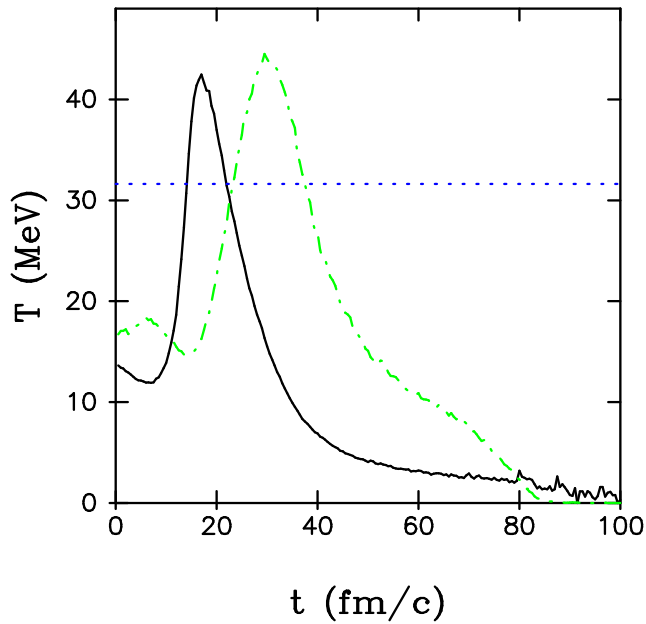}
\caption{
(Color online) Evolution of average temperature of the fireball (solid line) 
with the time. 
The total volume of the fireball is shown in arbitrary scale 
as dash-dotted line. 
The results were obtained using the BUU in reaction $^{48}$Ca+$^{48}$Ca 
at 400 AMeV using the impact parameter 1 fm. 
Dotted straight line shows the fireball temperature estimate for 
a given beam energy obtained from the 
systematics of pre-equilibrium spectra \cite{GelbkeBoal}. 
}
\label{fgtmptime}
\end{figure}

\subsection*{Implementation into the Boltzmann equation}

While the Boltzmann equation (\ref{BUU}) is formulated in terms of density, it does not 
explicitly consider temperature. Therefore temperature $T$ needs to be 
estimated independently. It is possible to estimate temperature using the 
Maxwellian momentum distribution of nucleons 

\begin{equation}
f(\vec{p})= \frac{1}{(2 \pi m T)^{3/2}} 
{\rm e}^{-\frac{p_x^2 + p_y^2 + p_z^2}{2 m T}},
\end{equation}
where $m$ is the nucleon mass. Using this formula, local temperature 
can be estimated from momentum distribution in the c.m. frame by 
evaluating the momentum variance. At early stages of collision this can be done 
primarily for transverse 
momentum since it provides a measure of mutual thermalization 
of particles from the projectile and target, which proceeds by 
distant elastic collisions generating the transverse momentum. 
More violent collisions would lead to emission of colliding nucleons and thus would not 
contribute to thermalization of the source. 
This temperature estimate can be done without requiring stopping and 
formation of the source equilibrated in all three dimensions, closer 
analogue would be the friction of two dilute gas clouds passing through 
each other. 

\begin{figure}[h]
\centering
\vspace{5mm}
\includegraphics[width=12.5cm,height=6.25cm]{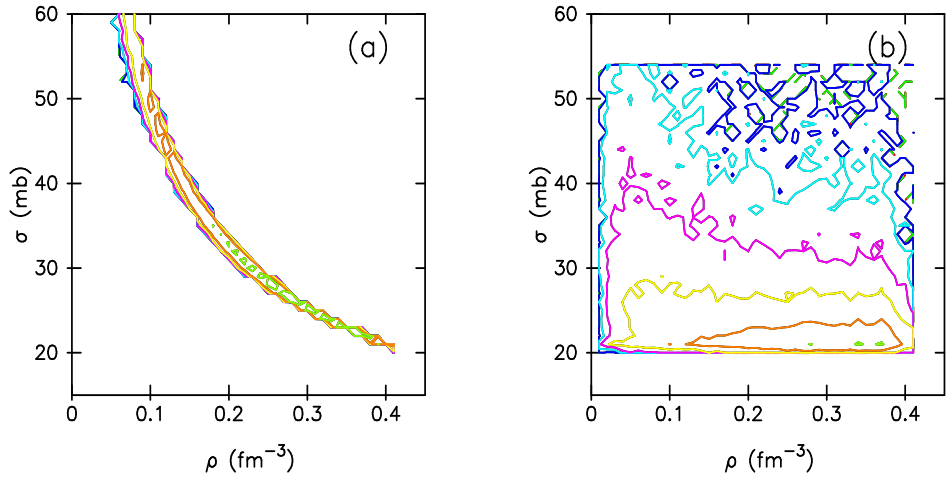}
\caption{
(Color online) Comparison of the nucleon-nucleon cross sections in two variants 
of the BUU calculations. 
On the left panel are the isospin-dependent nucleon-nucleon cross sections,
obtained as the proper volume of the Van der Waals form of the equation
of state, as a function of density, 
while on the right panel are shown the corresponding 
nucleon-nucleon cross sections, obtained using standard energy dependent 
parametrization, used in BUU calculation. 
The results were obtained using the BUU in reaction $^{48}$Ca+$^{48}$Ca 
at 400 AMeV using the impact parameter 1 fm. 
}
\label{fgcsrhoeos_old}
\end{figure}

Evolution of average temperature of the fireball 
with the time is shown in Fig. \ref{fgtmptime}. 
The results were obtained using the Boltzmann-Uehling-Uhlenbeck equation 
(BUU) \cite{BUU1,BUU2} in reaction $^{48}$Ca+$^{48}$Ca 
at 400 AMeV using the impact parameter 1 fm. 
Temperature was evaluated for each time step in the 
cubic cells with the side of 1 fm. 
Average temperature was determined as a mean value of temperature 
over all cells where number of nucleons was sufficient (corresponding to density of $\rho_0/10$) and temperature thus could be evaluated. 
The total volume of this fireball is shown in Fig. \ref{fgtmptime} 
in arbitrary scale as dash-dotted line. 
Dotted line shows the estimate of fireball temperature for 
a given beam energy obtained from the 
systematics of pre-equilibrium spectra \cite{GelbkeBoal}. 
One can see that the average temperature over the fireball 
at its peak value exceeds 
the estimate from the systematics, while the 
value averaged over the lifetime of the hot fireball (between 10 -- 30 fm/c) appears 
to correspond to the value from the systematics. Thus it appears 
that the procedure introduced here leads to reasonable estimate  
of local temperature. 

\begin{figure}[h]
\centering
\vspace{5mm}
\includegraphics[width=8.5cm,height=8.25cm]{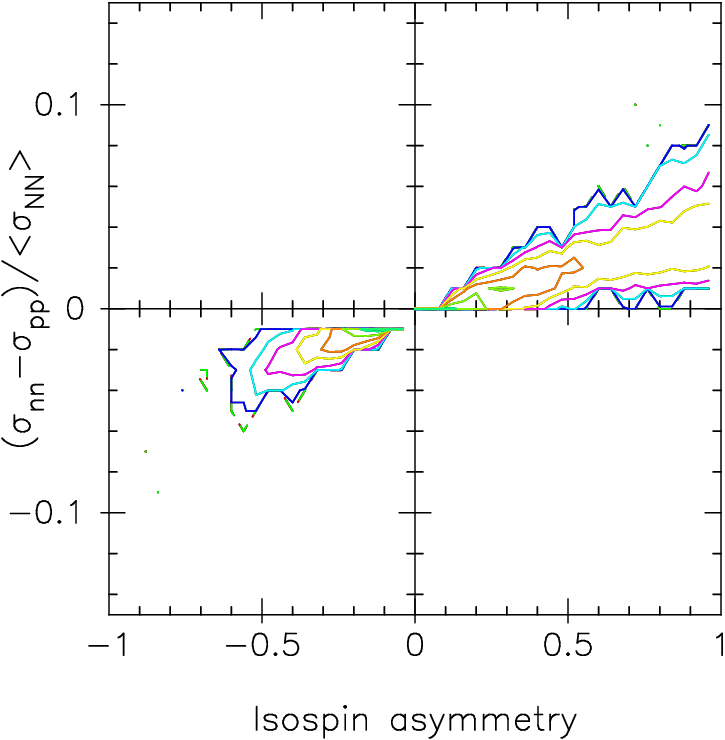}
\caption{
(Color online) Relative difference of the isospin-dependent neutron-neutron and 
proton-proton cross sections as a function of isospin asymmetry 
of the volume cell. 
The results were obtained using the BUU in reaction $^{48}$Ca+$^{48}$Ca 
at 400 AMeV using the impact parameter 1 fm. 
}
\label{fgdcsasym}
\end{figure}

Since the temperature, determined using the assumption of Maxwellian distribution 
represents the classical Boltzmann statistics, it can be corrected in order 
to reflect the Fermi statistics, which fermions like nucleons obey. 
To achieve this, one needs to multiply the classical temperature $T_{Boltz}$, corresponding 
to the Boltzmann statistics, by a factor 
$\langle\frac{f_{5/2}(z)}{f_{3/2}(z)}\rangle^{-1}$ 
and thus the formula (\ref{bfermi}) will turn into 

\begin{equation}
b' = \frac {b\kappa \rho^{\kappa} 
+ {2 \gamma a_s}(\frac{\rho}{\rho_0})^{\gamma}\tau_z {I}} 
{ 
\rho T_{Boltz} + b\kappa \rho^{1+\kappa} + 
{2 \gamma \rho_0 a_s}(\frac{\rho}{\rho_0})^{1+\gamma}\tau_z {I}} ,
\label{bboltz}
\end{equation}

which corresponds to classical case of Boltzmann 
statistics. Thus, remarkably, this classical expression can be used also 
for Fermionic (or even bosonic) particles, obeying their 
corresponding statistics. From practical point of view, in this way 
the non-trivial determination of the Fermionic temperature, depending of fugacity, 
can be avoided. 

Once the local temperature is determined, it is possible to implement the 
isospin-dependent nucleon-nucleon cross section, obtained using the 
formulas (\ref{bboltz}) and (\ref{csvdw}), to the reaction simulation which can be used 
for determination of collision rate. 
The results for solution 
of such fully isospin-dependent version of Boltzmann-Uehling-Uhlenbeck equation 
will be presented in the next section. 

\begin{figure}[h]
\centering
\vspace{5mm}
\includegraphics[width=14.5cm,height=9.75cm]{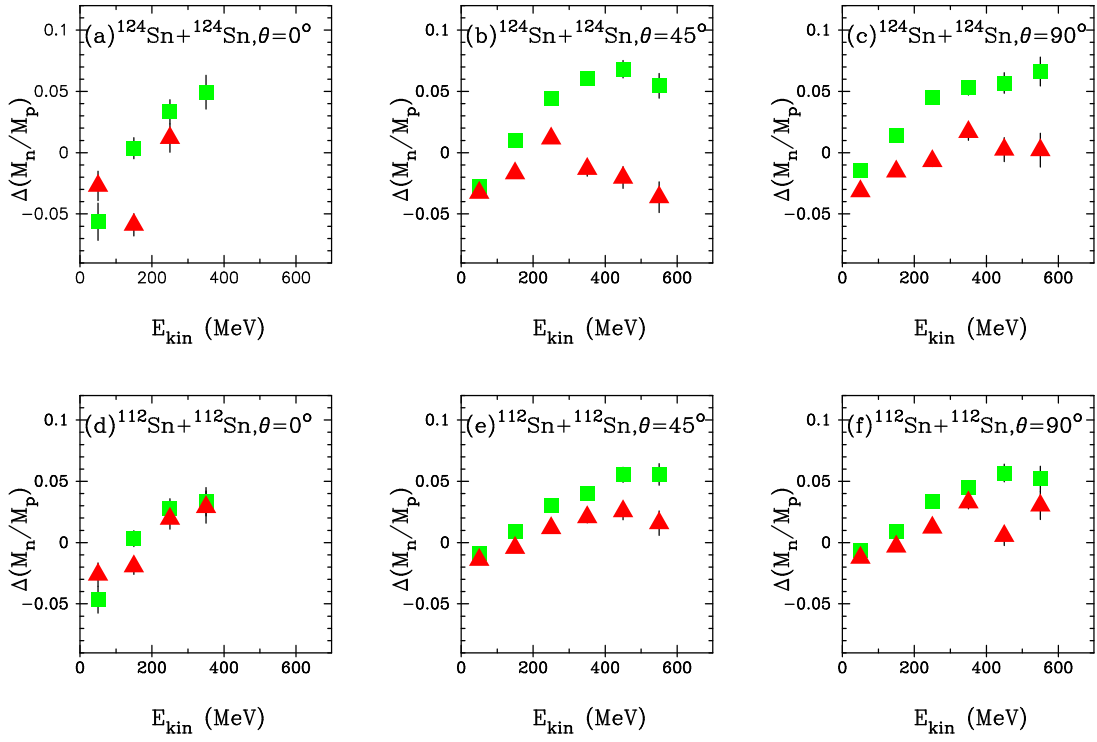}
\caption{
(Color online) Evolution of the difference of n/p multiplicity ratio between  
VdWBUU 
and 
fBUU calculations 
for  three angular ranges.  
The results were obtained in reactions $^{124}$Sn+$^{124}$Sn 
and $^{112}$Sn+$^{112}$Sn at 400 AMeV using the impact parameter 1 fm 
at the time 200 fm/c. 
Squares show the result with stiff symmetry energy while triangles 
show results for soft symmetry energy. Soft nuclear equation of state  
is used. 
}
\label{fgdnpthasft1fm}
\end{figure}

\section*{Reaction simulations}

The behavior of the in-medium nucleon-nucleon cross sections was investigated using the already mentioned reaction $^{48}$Ca+$^{48}$Ca 
at 400 AMeV using the impact parameter 1 fm. 
The soft equation of state was used, leading to incompressibility coefficient $K$=200 MeV. 
For the isospin asymmetric part the "asystiff" parametrization was used 
with two symmetry energy terms, the kinetic term with the parameters $a_{s1} $= 12.5 MeV and $\gamma_{1} = 2/3$, resulting from the Pauli principle, and the potential term with the parameters 
$a_{s2} $= 17.5 MeV and $\gamma_2$ = 2, respectively. 

\begin{figure}[h]
\centering
\vspace{5mm}
\includegraphics[width=14.5cm,height=9.75cm]{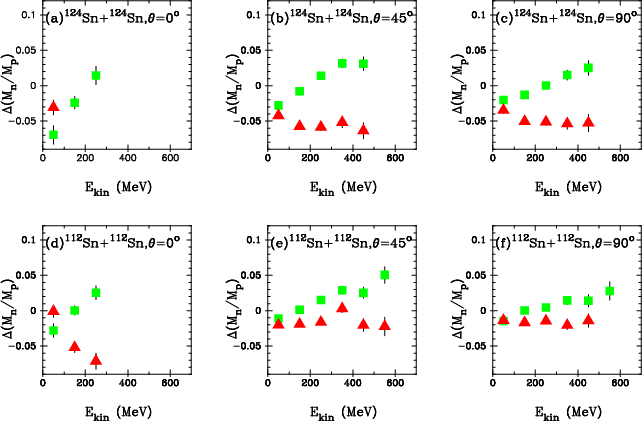}
\caption{
(Color online) Same as Figure \ref{fgdnpthasft1fm} but for stiff nuclear equation of state. 
}
\label{fgdnpthastf1fm}
\end{figure}

Fig. \ref{fgcsrhoeos_old} shows a 
comparison of the nucleon-nucleon cross sections in two variants 
of the Boltzmann-Uehling-Uhlenbeck simulations. 
On the left panel are shown, as a function of density, 
the isospin-dependent nucleon-nucleon cross sections, 
obtained using the proper volume of the Van der Waals-like equation 
of state. 
On the right panel are shown the corresponding free 
nucleon-nucleon cross sections, obtained as energy dependent
parametrization of measured nucleon-nucleon cross sections \cite{Cugnon}. 
It is apparent that while the isospin-dependent nucleon-nucleon cross sections 
essentially follow the $1/\rho^{2/3}$-dependence, the nucleon-nucleon cross section 
parametrization of Cugnon et al. 
leads to much larger spread, mostly due to its explicit energy dependence. 
Nevertheless, one observes that both parametrization 
cover essentially the same range of values of the nucleon-nucleon cross sections. 
Furthermore, from the comparison \cite{CassingMosel} of the parametrization of Cugnon et al. 
to in-medium cross sections at saturation density, calculated 
using the G-matrix theory by Cassing et al. \cite{Cassing}, it can be judged 
that the in-medium cross sections, obtained using the proper volume of the 
Van der Waals-like equation of state, are in better agreement with somewhat 
higher values of G-matrix in-medium cross sections of Cassing et al., which reflect properly 
the Fermionic nature of nucleons. 

In general, it is remarkable, that the in-medium nucleon-nucleon cross sections 
can be possibly directly related to the equation of state of the isospin 
asymmetric nuclear matter. This offers a more consistent description of the 
nuclear reactions and various astrophysical objects and processes in term of 
properties of nucleonic matter, expressed using the corresponding equation of state. 
However, one has to take into account that the equation of state of the isospin
asymmetric nuclear matter describes isotropic medium and thus 
the extracted in-medium cross sections represent angle-averaged values. These 
values are used in this work, and compared to the results obtained with angle-averaged 
free cross sections of Cugnon et al. \cite{Cugnon}, and thus the effect of 
the equation of state of the isospin
asymmetric nuclear matter on in-medium cross sections is demonstrated. 
However, one can consider possibility to implement 
angular dependence, either using the compatible microscopic calculations 
or from the observed experimental free nucleus-nucleus 
cross sections. This possibility is beyond the scope of the present work 
and will be investigated in our future work. 

\begin{figure}[h]
\centering
\vspace{5mm}
\includegraphics[width=14.5cm,height=9.75cm]{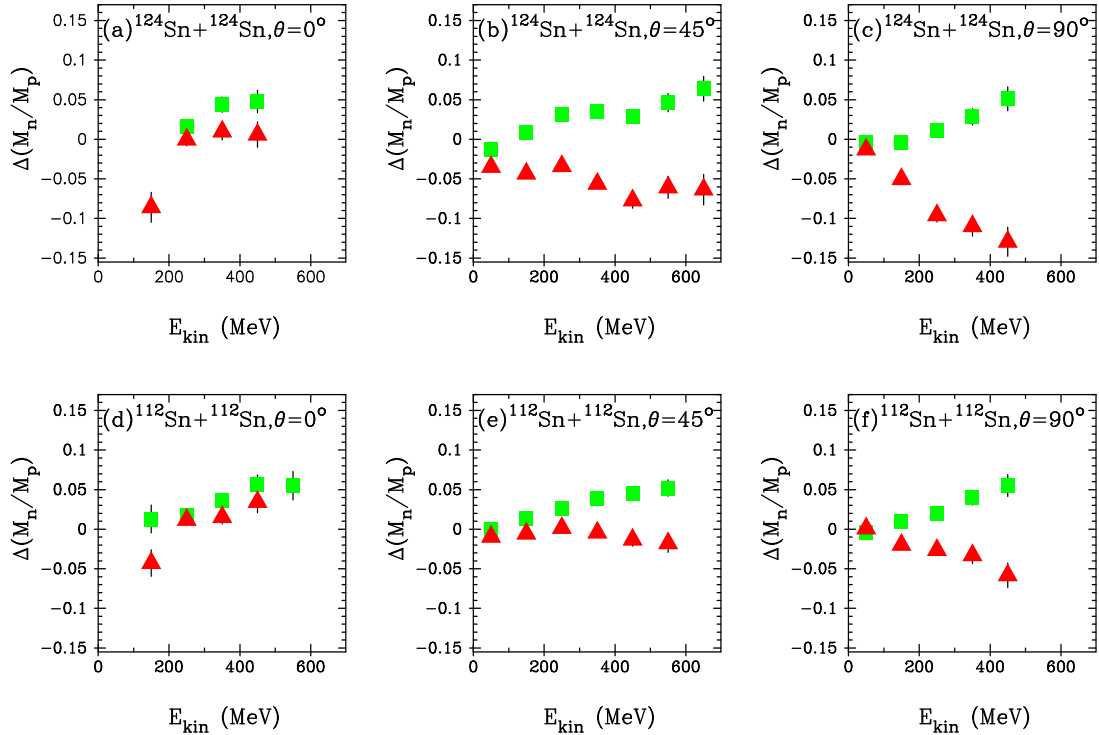}
\caption{
(Color online) Evolution of the difference of n/p multiplicity ratio between  
VdWBUU 
and 
fBUU calculations 
for  three angular ranges.  
The results were obtained in reactions $^{124}$Sn+$^{124}$Sn 
and $^{112}$Sn+$^{112}$Sn at 400 AMeV using the impact parameter 6 fm 
at the time 200 fm/c. 
Squares show the results with stiff symmetry energy while triangles 
show results for soft symmetry energy. Soft nuclear equation of state  
is used. 
}
\label{fgdnpthasft6fm}
\end{figure}

The magnitude of the effect of isospin asymmetry on the nucleon-nucleon 
cross sections can be judged from Fig. \ref{fgdcsasym}, which shows 
the relative difference of the isospin-dependent neutron-neutron and 
proton-proton cross sections as a function of isospin asymmetry 
of the volume cell. 
The results were again obtained using the BUU in reaction $^{48}$Ca+$^{48}$Ca 
at 400 AMeV using the impact parameter 1 fm. 
One can see that the relative magnitude does not reach very high 
values even for most isospin-asymmetric cells, and the sensitivity of the Boltzmann-Uehling-Uhlenbeck simulation to the isospin-dependent 
nucleon-nucleon cross sections will result from the cumulative effect of large amount of 
nucleus-nucleus collisions. 

The behavior observed in Figures \ref{fgcsrhoeos_old} and \ref{fgdcsasym} appears to be 
consistent with the behavior shown in Figure 4 of the work \cite{LiBAnn}. In both cases 
the absolute values of cross sections decrease monotonously with increasing density while the relative difference of the neutron-neutron and proton-proton cross sections increases with increasing asymmetry. Thus it appears that the procedure used to extract in-medium nucleon-nucleon cross sections by determining the parameters of corresponding Van der Waals EoS reflects the same physics, which is encoded even to a phenomenological equation of state such as Equation (\ref{EqPot}). 

\begin{figure}[h]
\centering
\vspace{5mm}
\includegraphics[width=14.5cm,height=9.75cm]{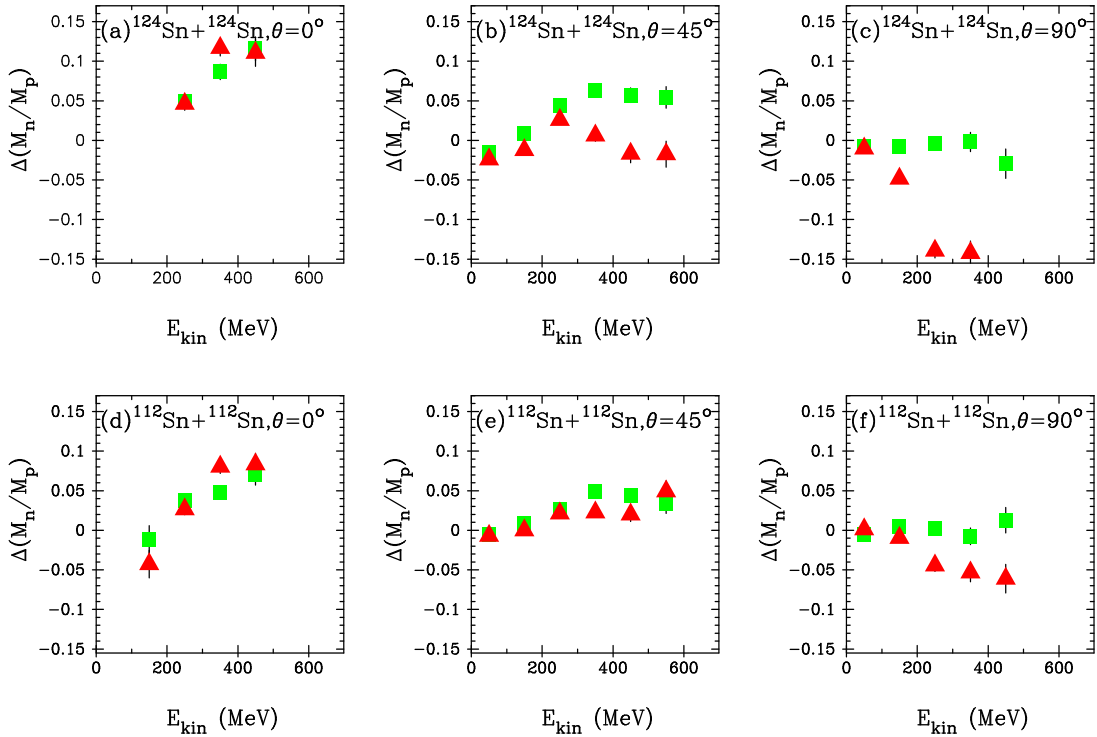}
\caption{
(Color online) Same as Figure \ref{fgdnpthasft6fm} but for stiff nuclear equation of state. 
}
\label{fgdnpthastf6fm}
\end{figure}

\begin{figure}[h]
\centering
\vspace{5mm}
\includegraphics[width=10.5cm,height=10.25cm]{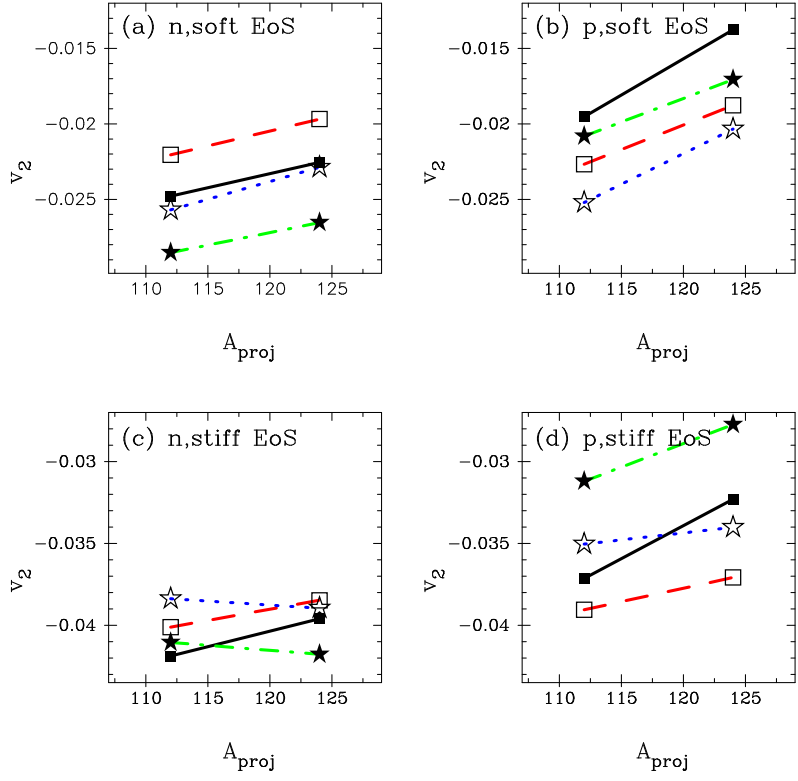}
\caption{
(Color online) Elliptic flow of neutrons and protons 
in reactions $^{124}$Sn+$^{124}$Sn 
and $^{112}$Sn+$^{112}$Sn at 400 AMeV at the impact parameter 6 fm. 
Solid and open squares show results of 
VdWBUU calculation 
and fBUU calculation, 
respectively, with soft symmetry energy.     
Solid and open asterisks show analogous results with stiff symmetry energy.  
}
\label{fgeflow6fm}
\end{figure}

Figure \ref{fgdnpthasft1fm} shows 
evolution of the difference of n/p multiplicity ratios between  
Boltzmann-Uehling-Uhlenbeck simulation with both 
isospin-dependent mean-field and nucleon-nucleon cross sections which are correlated to each other by the Eq.~\ref{csvdw} (based on the analogy with the van der Waals equation of state, 
thereafter we call this simulation as VdWBUU)
and Boltzmann-Uehling-Uhlenbeck simulation 
with isospin-dependent mean-field and free nucleon-nucleon cross sections (thereafter we call it fBUU), for 
three angular ranges as a function of the kinetic energy in the 
center-of-mass system.  
The results were obtained using the BUU simulation in reactions $^{124}$Sn+$^{124}$Sn 
and $^{112}$Sn+$^{112}$Sn at 400 AMeV using the impact parameter 1 fm 
at the stopping time 200 fm/c. 
Squares show the result with stiff symmetry energy parametrization 
($a_{s1} $= 12.5 MeV, $\gamma_{1} = 2/3$, 
$a_{s2}$ = 17.5 MeV and $\gamma_2 = 2$) 
while triangles 
show results for with soft symmetry energy parametrization 
($a_{s1}$ = 12.5 MeV, $\gamma_{1} = 2/3$, 
$a_{s2}$ = 17.5 MeV and $\gamma_2 = 1/2$). Soft nuclear equation of state 
($K$ = 200 MeV) is used in this case. Particles are considered as emitted when they are  
separated in the phase-space from any other particle and separation 
is large enough to assure that two particles are not part of a cluster ( a condition $\Delta \vec{p} \Delta \vec{r} > 2 h$ is implemented ). 
One can see that implementation of 
isospin-dependent nucleon-nucleon cross sections leads to 
significant variation of n/p multiplicity ratio and this 
effect appears to evolve with both kinetic energy and polar angle. 
Variation of n/p multiplicity ratio is more significant for the 
more neutron-rich system, which offers a strong argument for the 
use of neutron-rich exotic beams for studies of density dependence of 
the symmetry energy in the future. 

Figure \ref{fgdnpthastf1fm} again shows 
evolution of the difference of n/p multiplicity ratio between  
VdWBUU calculation 
and fBUU calculation,
in this case using stiff nuclear equation of state ($K$ = 380 MeV). 
Also in the case of stiff nuclear equation of state 
one can see that implementation of 
isospin-dependent nucleon-nucleon cross sections leads to 
considerable variation of n/p multiplicity ratio, which again 
evolves with both kinetic energy and polar angle. 
Also here the variation of n/p multiplicity ratio 
is more significant for the more neutron-rich system. 
It thus appears that variation of n/p multiplicity ratio between
VdWBUU calculation 
and fBUU calculation
provides 
a robust signal of the density dependence of nuclear symmetry energy. 

Figures \ref{fgdnpthasft6fm} and \ref{fgdnpthastf6fm} show results for 
difference of n/p multiplicity ratio between
VdWBUU calculation 
and fBUU calculation
analogous to figures \ref{fgdnpthasft1fm} and \ref{fgdnpthastf1fm}, with the impact parameter set to be 6 fm. It can be seen that the effect of isospin-dependent nucleon-nucleon cross sections persists, with comparable magnitude, even in peripheral collisions. This offers possibility to study such signal of the density-dependence of the symmetry energy in a wide range of centralities and thus eventually to provide a strong signal of the density-dependence of the symmetry energy in a wider range of nuclear density. 

The fact, that the isospin-dependent nucleon-nucleon cross sections,
obtained using the proper volume of the Van der Waals--like equation
of state, need to be introduced in order to fully explore the isospin dependence 
in the Boltzmann-Uehling-Uhlenbeck simulations, directly affects the applicability  
of the symmetry energy parametrizations 
to the study of astrophysical objects such as 
neutron stars or supernovae. 
Due to increased sensitivity to isospin due to isospin-dependent nucleon-nucleon cross sections,  the symmetry 
energy parametrizations may change significantly and that will affect 
the extrapolations toward the nuclear densities, typical for 
neutron stars and similar objects. On the other hand, 
increased sensitivity may offer more possibilities for the 
simulations of reactions of exotic nuclear beams, with the 
possible observation of stronger isospin-dependent signals. 

The recently performed simulations based on the Ultrarelativistic Quantum Molecular Dynamics model (UrQMD) suggest that one of the most promising probe of the strength of the symmetry energy at supra-saturation densities is the difference of the neutron and proton (or hydrogen) elliptic flows \cite{Xiao09, Rus11, Tra09a, Tra10}. The simulations were performed using both stiff ($\gamma$= 1.5) and soft ($\gamma$ =0.5) symmetry energy parametrizations. An inversion of the relative strengths of the elliptic flow for neutrons and protons is observed when the symmetry energy parametrization is changed from the stiff behavior to the soft behavior. 
Neutron and proton directed and elliptic flows were measured a decade ago in $^{197}$Au +$^{197}$Au collisions at beam energies from 400 to 800 AMeV using the LAND neutron detector and the FOPI Phase 1 forward wall \cite{Lei93, Lam94}. Comparison of predictions of the UrQMD model \cite{Li06b} provided a constraint on symmetry energy at supra-saturation densities from transverse momentum dependence of the neutron and hydrogen's elliptic flow parameter $v_2$ measured in the $^{197}$Au +$^{197}$Au system with FOPI+LAND, suggesting a value of $\gamma =0.9 \pm 0.4$ \cite{Rus11}, in agreement with findings at sub-saturation densities. However the statistics of these data set severe limits on the conclusions that can be drawn by comparison to transport model calculations. 
The promising results of re-analysis of FOPI+LAND experiment initiated proposal for a new experiment \cite{Lem09},  which is one of the first dedicated explorations of the symmetry energy at high densities. 
The experiment uses the LAND calorimeter for neutron and charged particle detection, and the  impact parameter is determined with a detection system with high effective granularity at forward angles consisting of several CsI rings of the CHIMERA multi-detector \cite{Pag04} and the ALADIN Time-Of-Flight wall \cite{Sch95}. In addition, flow of light fragments are measured with the Krakow telescope \cite{Krak} array positioned opposite from LAND. Uncertainties are expected to be reduced by a factor of 4-5, thus allowing to constrain theoretical calculations. Expected results of this experiment will provide a welcome testing-ground for the model calculations introduced in the present work. 

It is of interest to estimate what effect will the introduction the isospin-dependent nucleon-nucleon cross sections, obtained as the proper volume of the Van der Waals--like equation of state, have on the resulting elliptic flow of neutrons and protons. 
Figure \ref{fgeflow6fm} shows calculated values of the 
elliptic flow of neutrons and protons (determined conventionally as second Fourier coefficient of the invariant 
triple differential distribution $v_2$ relative to reaction plane) 
in reactions $^{124}$Sn+$^{124}$Sn 
and $^{112}$Sn+$^{112}$Sn at 400 AMeV at the impact parameter 6 fm.   
Solid and open squares show results of 
VdWBUU calculation 
and fBUU calculation, 
respectively, with soft symmetry energy parametrization.     
Solid and open asterisks show analogous results with stiff symmetry energy parametrization.  
One can see that introduction of isospin-dependent nucleon-nucleon 
cross sections, varying the ratio of neutron-neutron and proton-proton collision rates and thus modifying the in-plane/out-of-plane emission ratios, influences the resulting values of the elliptic flow, and since the effect appears to vary between neutrons and protons, it will strongly influence the differential elliptic flow, thus making it a strong signature of the nuclear equation of state, as suggested in the work \cite{Rus11}. However, since the effect of symmetry energy on such differential elliptic flow tends to vary also with isoscalar part of the nuclear equation of state, it appears necessary to study differential elliptic flow in combination with other observables, such as the evolution of n/p multiplicity ratio at different polar angles and kinetic energies. Such a study, carried out on sufficiently neutron-rich system, can provide a good sensitivity to both isoscalar and isovector part of the nuclear equation of state.

\section*{Conclusions}

The extension of the Boltzmann-Uehling-Uhlenbeck model of nucleus-nucleus collision is presented. 
The isospin-dependent 
nucleon-nucleon cross sections are estimated using the proper volume extracted from the equation 
of state of the nuclear matter transformed into the form of the Van der Waals equation of state.  
The results of such simulations demonstrate  
the dependence on symmetry energy which typically varies  
strongly from the results obtained using only the isospin-dependent mean-field. 
The evolution of the n/p multiplicity ratio with angle and kinetic energy, 
in combination with the elliptic flow of neutrons and protons, provides 
a suitable set of observables for determination of the density dependence 
of the symmetry energy. 
The model thus provides an environment for testing of equations of state, 
used for various applications in nuclear physics and astrophysics. 

\vspace{1.cm}
Acknowledgments: This work is supported
in part by the Chinese Academy of Sciences Fellowships for the
international scientists under the Grant No. 2011T2J13 (M.V.), and  by
NSFC of China under contract Nos. 11035009, 10979074, the Knowledge
Innovation Project of CAS under Grant No. KJCX2-EW-N01 (Y.G.M.). M.V. has been also supported by the Slovak Scientific Grant Agency under contract 2/0105/11, and by the Slovak Research and Development Agency under contract APVV-0177-11.

\end{document}